\documentclass[useAMS,usenatbib]{mn2e}
\usepackage[dvips]{graphicx}
\usepackage{amsmath}
\usepackage{amsfonts}
\usepackage{amssymb}

\title[On the detectability of habitable exomoons]{On the detectability of habitable exomoons with \emph{Kepler}-class photometry}
\author[Kipping, Fossey \& Campanella]{David M. Kipping$^{1,2}$\thanks{E-mail: d.kipping@ucl.ac.uk}\footnotemark[1], Stephen J. Fossey$^{1}$ \& Giammarco Campanella$^{1,3}$ \\
$^{1}$Department of Physics and Astronomy, University College London, 
Gower Street, London WC1E 6BT, UK \\
$^{2}$The HOLMES collaboration \\
$^{3}$Dipartimento di Fisica, Universit\'a di Roma "La Sapienza", Ple Aldo Moro 5, 00185 Rome, Italy}
\begin{document}

\date{Accepted TBA. Received ; in original form 2009 March 23rd}

\volume{TBA} \pagerange{TBA} \pubyear{TBA}

\maketitle

\label{firstpage}

\begin{abstract}
In this paper we investigate the detectability of a habitable-zone exomoon around various configurations of exoplanetary systems with the \emph{Kepler Mission} or photometry of approximately equal quality.  We calculate both the predicted transit timing signal amplitudes and the estimated uncertainty on such measurements in order to calculate the confidence in detecting such bodies across a broad spectrum of orbital arrangements.  The effects of stellar variability, instrument noise and photon noise are all accounted for in the analysis.  We validate our methodology by simulating synthetic lightcurves and performing a Monte Carlo analysis for several cases of interest.

We find that habitable-zone exomoons down to $0.2 M_{\oplus}$ may be detected and $\sim 25,000$ stars could be surveyed for habitable-zone exomoons within \emph{Kepler}'s field-of-view.  A Galactic Plane survey with \emph{Kepler}-class photometry could potentially survey over one million stars for habitable-zone exomoons.  In conclusion, we propose that habitable exomoons will be detectable should they exist in the local part of the galaxy.
\end{abstract}

\begin{keywords}
techniques: photometric --- planets and satellites: general --- planetary systems ---  occultations --- methods: analytical
\end{keywords}

\section{Introduction}

With the successful launch of \emph{Kepler}, a mission specifically designed to detect habitable Earth-like planets, a new age in extrasolar planetary science is about to commence.  \emph{Kepler} is a mission designed to detect the transit of an Earth across the Sun with its highly sensitive photometric camera; more details can be found in \citet{bas05} and \citet{koc07}, as well as on the mission website (http://www.kepler.nasa.gov/sci).

In this paper, we evaluate the range of exomoons that the \emph{Kepler Mission} or \emph{Kepler}-class photometry (KCP) could detect through transit timing effects, with particular attention to habitable-zone exomoons.  We find that habitable exomoons down to $0.2 M_{\oplus}$ should be detectable with the expected performance of \emph{Kepler}.  The theory behind detecting exomoons through timing effects was first proposed by \citet{sar99} and has been recently advanced by \citet{kipa09} and \citet{kipb09} and these models will be utilised throughout our analysis.

The underlying principle is that an exomoon should induce transit-time variations (TTV) and transit-duration variations (TDV) on the host planet.  These effects are predicted to exhibit a $\pi/2$ phase difference providing a unique exomoon signature.  In this work, we consider that both effects must be detected in order to claim a moon is present.  Additionally, measuring both timing effects allows for the determination of the exomoon's mass and orbital distance.

The question as to whether current telescopes can detect exomoons could have far-reaching implications for the future of exoplanet science.  Potentially, exomoons could be common habitable environments in the galaxy and would therefore be of great interest to astrobiologists.  In this paper, we find that exomoons are indeed already quite detectable with KCP.  We emphasise the use of \emph{Kepler}-class photometry due to the increasingly impressive results being obtained from the ground which are matching space-based photometry, for example \citet{joh09}.  Furthermore, ground-based observations are often more ideally suited for transit-timing studies due to the fewer constraints placed on the system, such as telemetry-limited data-download speeds.

\section{Transit Timing Effects due to an Exomoon}

\citet{sar99} predicted that an exomoon should induce transit time variations (TTV) on a host exoplanet by virtue of the fact that the planet and moon orbit a common centre of gravity.  More recently, the theory of exomoon detection has been refined in \citet{kipa09} and \citet{kipb09}.  According to these papers, there should exist two principal transit timing effects:

\begin{enumerate}
\item Transit time variation (TTV) caused by the position of the planet oscillating around the common centre-of-gravity of the planet-moon system.
\item Transit duration variation (TDV) caused by the apparent velocity of the planet increasing and decreasing as it moves around the centre-of-gravity of the planet-moon system.
\end{enumerate}

TTV and TDV must be detected in order to differentiate between a signal due to an exomoon or, for example, a perturbing planet (see \citet{ago05} and \citet{hol05}).  This is possible since TTV and TDV are predicted to be $\pi/2$ out-of-phase for an exomoon, thereby producing a unique detection signature. 

We briefly mention that other techniques for detecting exomoons have been proposed in the form of microlensing, \citet{han08}, lightcurve distortions, \citet{sza06}, planet-moon eclipses, \citet{cab07}, pulsar timing, \citet{lew08} and very weak distortions in the Rossiter-McLaughlin effect of a transiting planet, \citet{sim09}.

\section{Modelling the Detectability of Exomoons}
\subsection{Confidence of detection}

In order to explore a large range of parameter space, it is more convenient and efficient to employ analytic expressions rather than repeated individual simulations for thousands of different scenarios.  In order to proceed we need general expressions for the following:

\begin{enumerate}
\item TTV \& TDV signal amplitudes
\item Transit mid-time and transit-duration errors
\item Confidence of detection, based on signal-to-noise
\end{enumerate}

The TTV and TDV root mean square (r.m.s.) amplitudes may be easily calculated using the equations presented in \citet{kipa09} and \citet{kipb09}.  However, we also require expressions for the timing errors, which are critical in evaluating the signal-to-noise.

To address this, we will use the analytic expressions for the uncertainty on the mid-transit time ($t_{c}$) and duration ($T$) as derived by \citet{car08} using a Fisher-analysis of a trapezoid-approximated circular-orbit lightcurve.  These expressions consider the transit duration to be defined as the time taken for the planet to move between contact points 1.5 to 3.5 (see figure 1 of \citet{car08}).  This is in contrast to the definition used by \citet{sea03}, for example, who consider $t_T$ as being the duration between contact points 1 \& 4 and $t_F$ as the duration between contact points 2 \& 3 (see figure 1 of \citet{sea03}).

For the purposes of TDV measurements, the primary requirement is to use a measure of transit duration which has the lowest possible uncertainty.  By calculating the covariances of the lightcurve, \citet{car08} were able to show that $T$ can be calculated more precisely than either $t_T$ or $t_F$ and so we select $T$ as a robust duration parameter to explore the TDV effect.  The uncertainties on transit depth, $d$, transit duration, $T$, and mid-transit time, $t_C$, were derived by \citet{car08} to be:

\begin{align}
\sigma_d &= W^{-1} d \\ 
\sigma_T &= W^{-1} \sqrt{2 T \tau} \\
\sigma_{t_c} &= W^{-1} \sqrt{T \tau/2} \\
W &= d \sqrt{\Gamma_{ph} T}
\end{align}where $\tau$ is the ingress/egress duration, $\Gamma_{ph}$ is the photon collection rate, and $d$ is the transit depth\footnote{We choose to change the notation slightly from the original paper to avoid confusion with other parameters we use here}.  

These expressions do not hold for a poorly sampled ingress or egress and therefore we assume a cadence of 1 minute, corresponding to \emph{Kepler's} transit-timing/asteroseismology mode.  

The equations of \citet{car08} require the ingress duration\footnote{Defined as the duration between contact points 1 \& 2, which is of course equivalent to the egress duration for a circular orbit}, $\tau$, and the transit duration, $T$, as inputs.  In order to compute these values we will use the expressions of \citet{sea03} modified for the time between contact points 1.5 \& 3.5.  Note, that we do not require a more elaborate model, such as that of \citet{kip08}, since we are dealing with circular orbits only.  For the \emph{Kepler Mission} or KCP, we employ the same estimate for $\Gamma_{ph}$ as that of \citet{bor05} (B05) and \citet{yee08}:

\begin{equation}
\Gamma_{ph} = 6.3 \times 10^{8} \, \mathrm{hr}^{-1} \, 10^{-0.4 (m-12)},
\end{equation} where $m$ is the apparent magnitude.

For a normal transit depth observed $n$ times, the confidence, $C$, to which the transit is detected, in terms of the number of standard deviations, is defined by B05 as:

\begin{equation}
C(\mathrm{photometric}) = \frac{d}{\sigma_d} \sqrt{n}.
\end{equation}

The transit timing signals due to an exomoon are periodic in nature and so require a different detection method.  Typically, this problem is tackled by searching for significant peaks in a periodogram, as often employed for radial velocity searches (e.g. \citet{but02}).  However, this approach is less useful for transit timing effects due to an exomoon since the frequency we are trying to detect will always be much higher than the sampling frequency, as pointed out by \citet{kipa09}.  Since we are always below the Nyquist sampling rate, then only a range of possible harmonic frequencies are retrievable and a hence a set of possible exomoon masses.

We propose here that the ideal method is to search for statistically significant excess variance and then use the $\chi^2$-distribution to calculate the confidence of signal detection.  In order for this method to be applicable, we require a) that the uncertainty estimates are  robust and accurate; and b), that the period of the signal may be derived from amplitude information alone (which may then be compared to a periodogram to further refine the frequency).

The first of these requirements can be seen to be valid as several investigations have verified.  \citet{hol06} derived the uncertainties of the mid-transit times for four transits of XO-1b using three different methods: i) $\Delta \chi^2 = 1$ perturbation of the best-fit; ii) Monte Carlo bootstrapping; iii) Markov-Chain Monte Carlo (MCMC).  The authors found that all three methods produced very similar uncertainties, which implies the uncertainty estimates are highly robust.  Another example we point out is that of \citet{car08} who showed that the uncertainties derived using an MCMC-analysis were very similar to those predicted using analytic arguments.

The second of our requirements is validated by \citet{kipa09}, where it was shown that the ratio of the TTV and TDV signal amplitudes may be used to obtain the period of the exomoon.  This period may then be compared to the set of possible harmonic frequencies derived from a periodogram in order to obtain a highly reliable estimate.  We therefore conclude that a search for excess variance is the most appropriate strategy for searching for exomoons through transit timing effects.  The confidence of detection may be found by integrating the probability density function of the $\chi^2$-distribution.

\begin{equation}
C(\mathrm{timing}) = \sqrt{2} \mathrm{erf}^{-1}\Big[1-\int_{\alpha^2}^{\infty} \frac{x^{(n/2)-1} \exp^{-x/2}}{2^{n/2} \Gamma(n/2)}  \mathrm{d}x\Big] ,
\end{equation}where $n$ is the number of transits observed and $\alpha^2$ is the observed value of $\chi^2$, given by:

\begin{equation}
\alpha^2 = n\Big(1 + \frac{\delta^2}{\Delta^2}\Big),
\end{equation} where $\delta$ is the r.m.s. amplitude of the transit timing signal (see \citet{kipa09}) and $\Delta$ is the uncertainty on the mid-transit time/transit duration.

Integrating and making the above substitution we have:

\begin{equation}
C(\mathrm{timing}) = \sqrt{2} \mathrm{erf}^{-1}\Big[1-\mathrm{Q}\Big\{\frac{n}{2},\frac{n}{2} \Big(1+\frac{\delta^2}{\Delta^2}\Big)\Big\}\Big],
\end{equation} where $\mathrm{Q}\{a,b\}$ is the incomplete upper regularized Gamma function.  We summarise our assumptions below:

\begin{itemize}
\item[{\tiny$\blacksquare$}]  Only one exomoon exists around the gas giant exoplanet of interest.
\item[{\tiny$\blacksquare$}]  The moon and planet are both on circular orbits and the moon's orbit is prograde.
\item[{\tiny$\blacksquare$}]  The moon's orbital plane is coaligned to that of the planet-star plane which is itself perpendicular to the line-of-sight of the observer, i.e. $i = 90^{\circ}$.
\item[{\tiny$\blacksquare$}]  If a planet is within the habitable zone, then any moon around that planet may also be considered to be habitable.
\item[{\tiny$\blacksquare$}]  A transiting planet must be detected to 8-$\sigma$ confidence to be accepted as genuine.
\item[{\tiny$\blacksquare$}]  An exomoon must be detected through either a) TTV to 8-$\sigma$ and TDV to 3-$\sigma$ confidence or b) TTV to 3-$\sigma$ and TDV to 8-$\sigma$ confidence, in order to be accepted as genuine.
\item[{\tiny$\blacksquare$}]  The \emph{Kepler Mission} or KCP will be used in high cadence mode for the transit timing of a target of interest for $\simeq 4$ years.
\item[{\tiny$\blacksquare$}] $n = M/P_P$ where $M$ is the mission duration and $P_P$ is the period of transiting planet.  
\item[{\tiny$\blacksquare$}]  At least three transits are needed to detect both a planet and a moon.
\end{itemize}

In most of the cases we will consider, many more than three transits will be detected and three can be seen to be the limiting case for G0V stars, where the habitable zone is sufficiently distant to only permit three transits in a 4-year timespan.  Although statistically speaking three transits is sufficient, there is a risk of an outlier producing a false positive.  We therefore consider detections of habitable exomoons in early G-type star systems to be described as `tentative', whereas once four transits are detected, for stars of spectral type G5V and later, this risk can be considered to be reduced.

The nominal mission length of \emph{Kepler} is 3.5 years and it may be extended to up to 6 years, which justifies our choice of 4 years of transit timing observations.  A ground-based search achieving KCP may easily be operational for 4 years or more.  We choose 8-$\sigma$ as the signal detection threshold since this is the same as that used by \emph{Kepler}.  The second signal may be detected to lower significance since it is only used to confirm the phase difference between the two and also derive the exomoon period.

\subsection{The total noise}

The expressions of \citet{car08} only consider shot noise through the $\Gamma_{ph}$ parameter.   However, if we assume that the impact on $\sigma_T$, $\sigma_{t_c}$ and $\sigma_d$ are approximately equivalent for additional uncorrelated noise and for correlated noise, then we may simply modify $W$ to absorb the effects of red noise.  For uncorrelated noise we may add the additional sources of noise in quadrature.  Note, that this is the same treatment utilised in the design technical documents for \emph{Kepler}, for example see B05.

In general, there are expected to be three major types of noise present in the \emph{Kepler} data in the form of shot noise, instrument noise and stellar variability.  Instrument noise is due to a variety of effects and has been modelled in depth by B05 (and \citet{koc04}) to quantify its effect as a function of magnitude.  With all three noise sources, we modify $W$ to $W'$, given by:

\begin{equation}
\frac{1}{W'} = \frac{1}{d} \sqrt{\frac{1}{\Gamma_{ph} T} + I^2 + S^2},
\end{equation} where $I$ is the instrument noise and $S$ is the stellar variability.

$I$ is a function of magnitude which may be calculated using the model of B05 and we show all three noise sources plotted as a function of magnitude in figure 1.  We assume a constant value for stellar variability of 10 ppm across all spectral types, a reasonable assumption, given that 65--70\% of F7-K9 main-sequence stars in the \emph{Kepler} field are likely to have similar or lower intrinsic variability than the Sun (\citet{bat02}; B05) on timescales important to transit detections.  We also note that this equation is equivalent to the formulation used in the original technical design papers for \emph{Kepler}, for example see equation (1) of B05.

\begin{figure}
\begin{center}
\includegraphics[width=8.4 cm]{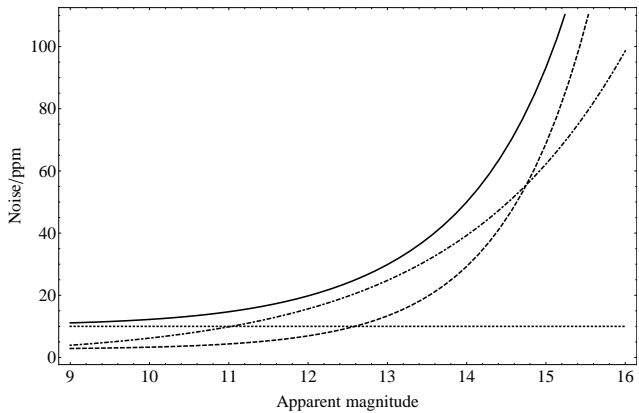}
\caption{\emph{Noise sources predicted to affect \emph{Kepler} photometry as a function of magnitude.  Instrument noise is dashed, photon noise is dot-dashed, stellar variability is dotted and the total is solid.  Values obtained from Bill Borucki in personal communication.}} \label{fig:fig1}
\end{center}
\end{figure}

\subsection{The habitable zone}

We choose to consider a moon-hosting gas-giant exoplanet around a variety of main-sequence stars as shown in table 1.  For each star we calculate the habitable zone orbital distance, $a_{hab}$, to be defined as the distance where a planet would receive the same insolation as the Earth.  This straightforwardly permits a reasonable estimate of the habitable zone for each star type.  For a more in-depth consideration of habitability of exomoons see \citet{wil97}.

\begin{equation}
a_{hab} = \sqrt{L_*/L_{\odot}} \, \mathrm{AU}.
\end{equation}

For each planet-moon system we consider, the period of the transiting planet is calculated using Kepler's Third Law:

\begin{equation}
P_{hab} = 2 \pi \sqrt{\frac{a_{hab}^3}{G (M_*+M_P + M_S)}}.
\end{equation}

We choose to work in the time domain, rather than the orbital-distance formulation, since a major limiting factor in our study is the \emph{Kepler Mission} duration.

\subsection{Properties of host star}

In our analysis, we will only consider single main-sequence stars which offer the best potential for hosting habitable environments.  We will consider spectral types from M5V to F0V and assume for each an approximate mass, radius, and effective temperature as given by \citet{cox00}, and a luminosity derived from data therein.  We use the \emph{Kepler} bandpass to calculate the absolute magnitude of these stars\footnote{Guidelines available from http://keplergo.arc.nasa.gov/proposal.html}.

For each stellar type, we assume the stars are not young and may be considered to be slow rotators.  Since stellar variability is correlated to rotational period (see \citet{dor94}), we therefore limit ourselves to quiet stars.  This is the same assumption used for \emph{Kepler}'s ability to detect Earth-like planets since very active stars will be too variable for the detection of such bodies.  In table 1 we list the different star properties.

\begin{table*}
\caption{\emph{Properties of stars used in our calculations.  Values taken from \citet{cox00}.  Absolute magnitudes in the \emph{Kepler} bandpass calculated using guidelines on the mission website.}} 
\centering 
\begin{tabular}{c c c c c c} 
\hline\hline 
Star type & $M_*$/$M_{\odot}$ & $R_*$/$R_{\odot}$ & $L_*$/$L_{\odot}$ & $T_{eff}$/K & $M_{\mathrm{Kep}}$ \\ [0.5ex] 
\hline 
M5V & 0.21 & 0.27 & 0.0066 & 3170 & 11.84 \\
M2V & 0.40 & 0.50 & 0.0345 & 3520 & 9.49 \\
M0V & 0.51 & 0.60 & 0.0703 & 3840 & 8.42 \\
K5V & 0.67 & 0.72 & 0.1760 & 4410 & 7.06 \\
K0V & 0.79 & 0.85 & 0.4563 & 5150 & 5.78 \\
G5V & 0.92 & 0.92 & 0.7262 & 5560 & 5.02 \\
G2V & 1.00 & 1.00 & 1.0000 & 5790 & 4.63 \\
G0V & 1.05 & 1.10 & 1.3525 & 5940 & 4.34  \\
F5V & 1.4 & 1.3 & 2.9674 & 6650 & 3.47 \\
F0V & 1.6 & 1.5 & 5.7369 & 7300 & 2.71 \\ [1ex]
\hline\hline 
\end{tabular}
\label{table:nonlin} 
\end{table*}

\subsection{Properties of exomoons}

Although no exomoons have yet been discovered, it is possible to calculate the range of exomoons which are dynamically stable around each planet.  \citet{bar02} addressed this problem and developed a set of analytic expressions, which can be shown to provide excellent agreement to numerical simulations, for the stability of exomoons around exoplanets.  Assuming an exomoon has to be stable for at least 5 Gyr, we are able to calculate the maximum allowed exomoon mass in each case example (see equations (7) to (9) of \citet{bar02}).  We assume Jupiter-like values for the tidal dissipation factor, $Q_P = 10^5$, and for the tidal Love number, $k_{2p} = 0.51$, as used by \citet{bar02}.

We also are able to estimate the range of allowed values for the planet-moon separation, in units of Hill radii, which we label as $\xi$.  \citet{dom06} presented the relevant expressions, which again can be shown to provide excellent agreement to numerical simulations.  Using their equation (5), we are able to estimate the maximum distance at which an exomoon is stable for prograde orbits.

For the minimum distance, we calculate the Roche limit of the planet in all cases.  If this value is greater than $2 R_P$ we use the Roche limit as the minimum distance, otherwise $2 R_P$ is adopted.  We assume that there is no reason for a moon to exist at any particular value of $\xi$ and thus the prior distribution is flat.

\section{Lightcurve Simulations}
\subsection{Lightcurve generation}

A critical assumption in this paper is the use of the equations of \citet{car08} for the uncertainties on the transit duration and mid-transit time.  The authors tested their expressions using synthetic lightcurves and an MCMC fitting procedure as well as a numerical Fisher-analysis.  They report excellent agreement between their expressions and the derived errors but find the greatest departure for poorly sampled ingresses and near-grazing transits.  Our choice of assumptions avoids both of these issues.

We note that these tests were run for shot noise only and in this paper we consider the effects of both instrument and stellar noise.  Having modified $W$ to $W'$ to account for these additional noise sources, we choose to test this modified formulation through generation of synthetic lightcurves.

In this work, we will use the \citet{man02} code to generate lightcurves accounting for limb darkening based upon a non-linear law from \citet{cla04}.  We use the \citet{sea03} expressions to compute the motion of the planet since we are dealing with circular orbits only.  We generate the lightcurves for a Neptune, Saturn and Jupiter-like planet in the habitable zone of a $m_{\mathrm{Kep}}=12$, G2V Sun-like star with $i = 90^{\circ}$ and $e = 0$.  Each lightcurve is generated to have 1000 data points evenly spaced with a 1-minute cadence centred on the mid-transit time.

The first noise source we add is shot noise generated by taking a random real number from a normal distribution of mean zero and standard deviation given by:

\begin{equation}
\sigma_{shot} = \frac{1}{\sqrt{t_{exp} \Gamma_{ph}}},
\end{equation} where $t_{exp}$ is the time between each consecutive measurement, which is assumed to be one minute in our calculations.

For the stellar noise, we generate 1000 sinusoidal waveforms with varying periods randomly selected between one minute up to 24 hours.  These waveforms are then coadded to give a single synthetic stellar signal designed to mimic the non-stochastic nature of real stellar variability.  The r.m.s. amplitude of the combined signal is then scaled to $10$ ppm, which matches the amplitude prediction of B05 for a G2V star.

We consider that the instrument noise is composed of hundreds of different noise sources periodic in nature varying on timescales from one minute to one day.  There is no benefit of including timescales longer than this since transit events will not last longer than $\sim 1 $ day.  The combined r.m.s. amplitude of the instrument noise is set to be given by the values calculated by Bill Borucki (personal communication) and is a function of visual magnitude.  For a $m_{\mathrm{Kep}}=12$ star, we take an r.m.s. instrument noise of 7.47 ppm.

\subsection{Lightcurve fitting}

The noisy lightcurves are generated 10,000 times with the correlated noises and photon noise being randomly generated in all cases.  The lightcurves are then passed onto a lightcurve fitting code used to obtain best-fit values for $T$ and $t_c$.  In all cases we fit for $a/R_*$, $i$, $R_P/R_*$ and $t_c$ and therefore assume that the out-of-transit baseline is well known and its errors are essentially negligible.  This is a reasonable assumption for high quality photometry with large amounts of out-of-transit data.

The fitting code finds the best-fit to the lightcurve by utilising a genetic algorithm\footnote{Available from http://whitedwarf.org/parallel/}, {\tt PIKAIA} (see \citet{met03}), to get close to a minimum in $\chi^2$.  This approximate solution is then used as a starting point for a $\chi^2$-minimisation performed with the AMOEBA routine (\citet{pre92}).  The AMOEBA solution is tested by randomly perturbing it and refitting in 20 trials.

In each subsequent fitting run of the 10,000 lightcurves, the AMOEBA routine starts from the original best-fit.  The resulting values of $t_C$ and $T$ are binned and plotted as a histogram (figure 2) and then compared to the distribution expected from the modified \citet{car08} expressions.

\subsection{Comparison to the analytic expressions}

For the three cases of a Neptune, a Saturn and a Jupiter, we obtain 10,000 estimates of $T$ and $t_c$ in each case.  We compare the distribution of $t_C$ and $T$ to the predicted distribution from the expressions of \citet{car08}.  In all cases, we find excellent agreement between the predicted uncertainties and the theoretical values.  In figure 2, we plot a histogram of the results for the Saturn-case $T$ values, and overlay the predicted value of $\sigma_{T}$ for comparison where the quality of the agreement is evident from the plot.  Note that this overlaid Gaussian is not a fit but a theoretical prediction.

\begin{figure}
\begin{center}
\includegraphics[width=8.4 cm]{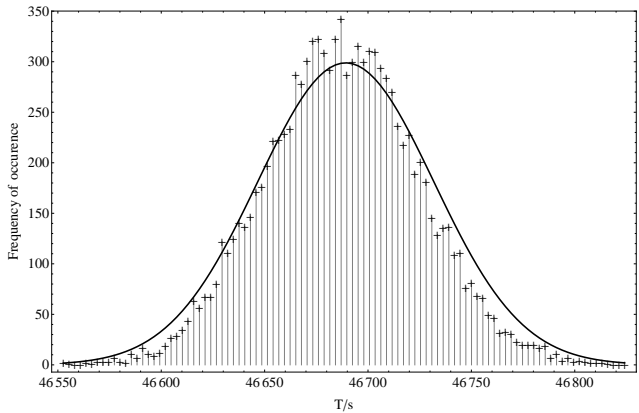}
\caption{\emph{Comparison of the distribution in $T$ found using Monte Carlo simulations (scattered points) of a transit of a Saturn-like planet versus that from theoretical prediction (smooth line).  The theoretical prediction for the timing error (42.6s) is slightly larger than that found in our simulations (37.3 s).}} \label{fig:fig2}
\end{center}
\end{figure}

We find excellent agreement for both $\sigma_T$ and $\sigma_{t_C}$, and if anything the theoretical expressions slightly overestimate the uncertainties.  Thus any results from this study can infact be considered to be slightly conservative.  We prefer to adopt a conservative approach in our analysis since there may be unexpected sources of noise which increase the timing uncertainties.

\section{Implementation of expressions}
\subsection{Earth-transit magnitude limit}

We first consider the magnitude limit in detecting an Earth-like transit which will allow us to compare the efficiency of detecting Earth-like planets and moons.  \emph{Kepler} is designed to look at $\sim 10^5$ stars between $6^\mathrm{th}$ to $16^\mathrm{th}$ magnitude\footnote{See http://kepler.nasa.gov/sci}.  For each spectral type in table 1, we are able to compute the faintest visual magnitude to which a habitable Earth-like transit can be detected.  \emph{Kepler} was conceived with the goal of detecting an Earth-Sun transit of 6.5 hours in duration, which represents about half the duration if the Earth has an impact parameter of zero.  However, the maximum magnitude to which an Earth-like transit could be detected will be for the cases of equatorial transits as these maximise the transit duration and hence integration time.  We establish the following criteria for a reliable transit detection:

\begin{itemize}
\item Each transit must be detected to $\geq$ 1-$\sigma$ confidence.
\item The folded lightcurve must have a significance of $\geq$ 8-$\sigma$.
\item The time between contact points 2 \& 3 $\geq$ 1 hour, in order to be detected with \emph{Kepler's} survey-mode cadence of 30 minutes.
\item At least 3 transits must be detected over the mission duration.
\end{itemize}

With these criteria, we compute the magnitudes of the faintest stars hosting a habitable, transiting Earth which can be detected with KCP, for both full transit durations (i.e. equatorial transits) and half-durations (i.e. impact parameter chosen to be such that the transit duration is half the full transit). The results are given in Table 2.  As expected, smaller host stars can be fainter due to their smaller radius and hence deeper transit depths.  We also pick up more transits towards smaller stars since the orbital period of the habitable Earth decreases.

\begin{table*}
\caption{\emph{Faintest stars for which a habitable transiting Earth could be detected for different star types.  Final column gives maximum distance of such a star with a full transit duration ($D_{full}$), based on absolute magnitude in the \emph{Kepler} bandpass.  A blank indicates that no magnitude can satisfy the detection criteria.}} 
\centering 
\begin{tabular}{c c c c c} 
\hline\hline 
Star type & $D_{half}$ $m_{\mathrm{Kep},\mathrm{max}}$ & $D_{full}$ $m_{\mathrm{Kep},\mathrm{max}}$ & $d_{min}$/pc & $d_{max}$/pc \\ [0.5ex] 
\hline 
M5V & - & 18.111 & 0.68 & 179.56 \\
M2V & 15.992 & 16.190 & 2.00 & 218.78 \\
M0V & 15.206 & 15.465 & 3.28 & 256.45 \\
K5V & 14.278 & 14.618 & 6.14 & 324.79 \\
K0V & 13.286 & 13.712 & 11.07 & 385.83 \\
G5V & 12.762 & 13.240 & 15.70 & 440.56 \\
G2V & 12.236 & 12.763 & 18.79 & 423.25 \\
G0V & 11.520 & 12.106 & 21.48 & 357.44 \\
F5V & - & - & - & - \\
F0V & - & - & - & - \\ [1ex]
\hline\hline 
\end{tabular}
\label{table:faint} 
\end{table*}

Even if \emph{Kepler} was extended to a mission length of 6 years, an F5V star would have a habitable zone so far out that detecting three transits within this region would be impossible.  Hence, we do not consider F type stars in our analysis.  We also consider $6^{\mathrm{th}}$-magnitude stars to be the bright limit, giving us $d_{min}$, the minimum distance for each spectral type (see Table 2), since the \emph{Kepler} field is specifically chosen to avoid such bright targets.

\subsection{Jupiters vs Saturns vs Neptunes}

We may use the approximate analytic expressions to get a handle on the general trends in exomoon detection.  The first question we may ask is what is the optimum planet to search for moons around, out of the three classes of Neptunes, Saturns and Jupiters?  We take each of these planets and simulate the detectability of the TTV signal as a function of planetary orbital period, $P_P$, for an $0.2 M_{\oplus}$ exomoon with $a_S = 0.4895$ Hill radii.  We fix the star to be a G2V, $m_{\mathrm{Kep}}=12$ object and work with $i = 90^{\circ}$ for simplicity.

We calculate the signal amplitude and mid-transit time uncertainty in all cases and hence find the $\chi^2$ value for a range of orbital periods.  In figure 3, we plot $\chi^2$ as a function of host planet's orbital period.  The plot reveals that Jupiters are the hardest to search for exomoons around whilst Saturns are the easiest.  This is due to Saturn's low density meaning a large transit depth but a low enough mass such that an exomoon still affects it significantly.  We find the same order of detectability consistently in many different orbital configurations and for TDV as well.

\begin{figure}
\begin{center}
\includegraphics[width=8.4 cm]{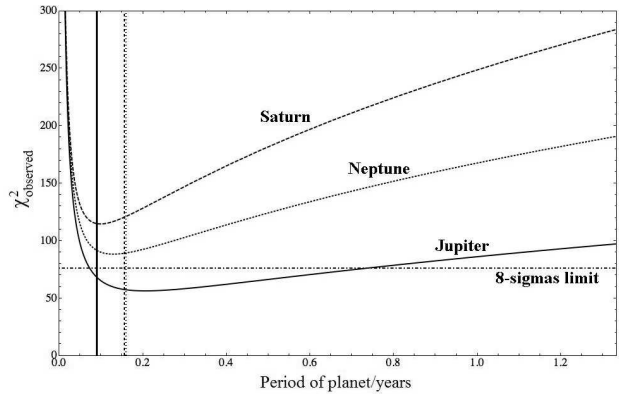}
\caption{\emph{Example plot showing the TTV detectability of a $0.2 M_{\oplus}$ ($\xi=0.4895$) exomoon around a Jupiter, Saturn and Neptune-like exoplanet for a G2V, $m_{\mathrm{Kep}}=12$ star.  Values of $\chi^2 \gg 76$ are detectable at $\geq$ 8-$\sigma$ confidence.  Saturns present the strongest signal due to their low density.  The vertical lines represent the stability limit of such a moon, calculated using the model of \citet{bar02}.}} \label{fig:fig3}
\end{center}
\end{figure}

\subsection{$\xi$-$m_{\mathrm{Kep}}$ parameter space}

We now consider an exomoon in the optimum condition of a Saturn hosting a single satellite in the habitable zone of the host star.  The detectability of the timing signals depends on the planet-moon separation, $a_S = \xi R_H$, the apparent magnitude of the host star in the \emph{Kepler} bandpass, $m_{\mathrm{Kep}}$, mass of the exomoon, $M_S$, and finally the host star's mass and radius (i.e. spectral type).  Note that $R_H$ denotes the Hill radius\footnote{Note that we denote the quantity $a_S/R_H$ as $\xi$, as opposed to $\chi$ in \citet{kipa09} and \citet{kipb09} to avoid confusion with the chi square distribution.}.

TDV increases with lower values of $\xi$ and TTV increases with higher values of $\xi$, but we must maintain a balance such that at least one of the effects is detected to 8-$\sigma$ and the other to 3-$\sigma$ confidence (dictated by our previous detection criteria).  To find the limits of interest, we assume a zero-impact-parameter transit of our optimum planet-moon arrangement.  The orbital period of the host planet is always given by $P_{hab}$.  We prefer to calculate this best-case scenario since we may then consider it unfeasible to detect habitable moons beyond this limit.

For a given star type and exomoon mass, we plot the contours of $m_{\mathrm{Kep}}$ and $\xi$ which provide TTVs of 3- and 8-$\sigma$ confidence and then repeat for TDVs of the same confidences.  There are two possible acceptance criteria: i) TTV confidence is $\geq 3$-$\sigma$ and TDV confidence is $\geq 8$-$\sigma$; ii) TTV confidence is $\geq 8$-$\sigma$ and TDV confidence is $\geq 3$-$\sigma$.  The loci of points below these lines represents the potential detection parameter space.  For each star type and exomoon mass, these loci will be different.  In figure 4 we show a typical example for an M0V-type star and $M_S = \frac{1}{3} M_{\oplus}$.

\begin{figure}
\begin{center}
\includegraphics[width=8.4 cm]{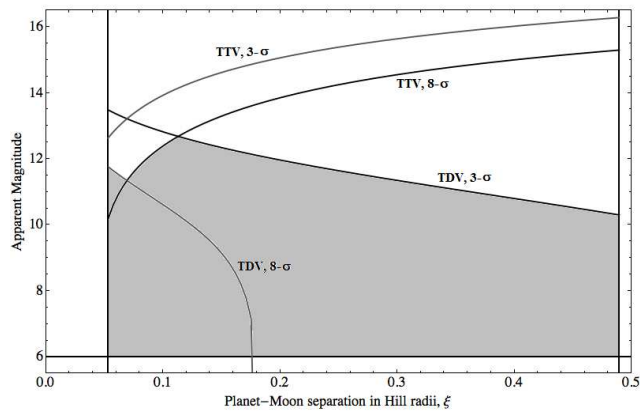}
\caption{\emph{Detectable range of a habitable $\frac{1}{3} M_{\oplus}$ exomoon for an M0V star with respect to apparent magnitude and $\xi$ parameter space, with \emph{Kepler}-class photometry.  The grey area represents the area satisfying our detection criteria of detecting TTV or TDV to 8-$\sigma$ and the other timing effect to 3-$\sigma$, shown by the four curves. Additional constraints are the lower Roche limit, the upper dynamical stability limit and the bright-star magnitude limit.}} \label{fig:fig4}
\end{center}
\end{figure}
There are certain cases which constrain the stars of interest.  The \citet{bar02} limit may be calculated for a Saturn harbouring an exomoon in a system of 5 Gyr age.  This suggests that the maximum moon mass that a habitable zone Saturn could hold around a M5V star would be $\sim 0.3$ Ganymede masses, assuming the maximum prograde orbital distance is 0.4895 Hill radii, as calculated by \citet{dom06}.  In contrast, an M2V star allows for a habitable Saturn to hold onto a $0.4$-Earth-mass moon for over 5 Gyr.

A second lower limit exists from tidal forces and the Roche limit.  An upper limit is given by the fact we require three transits in a 4 year observation duration and thus the most distant habitable zone assumed here corresponds to a period of 1.33 years, which excludes the F0V and F5V stellar types.

\subsection{Magnitude quartiles}

For figure 4, we are able to convert the plot into three key numbers by assuming that the probability distribution of exomoons with respect to $\xi$ is approximately flat.  This simple a-priori approximation allows us get a handle on some typical magnitude limits.  This assumption effectively converts the two-dimensional array of $\xi_i$-$m_{\mathrm{Kep},i}$ into a one-dimensional list of just $m_{\mathrm{Kep},i}$.  We then simply take the quartiles of this list to obtain estimates for the 25\%, 50\% and 75\% catch-rate values of $m_{\mathrm{Kep}}$.  Physically speaking, we are saying that, for example, at the 50\% catch-rate value of $m_{\mathrm{Kep}}$, there is a 50\% probability of detecting an exomoon if an exomoon exists with a flat prior distribution between $\xi_{min}$ and $\xi_{max}$.

Accordingly, we can calculate the magnitude limits to detect 25\%, 50\% or 75\% of the exomoons in the given sample, which we label as the upper, middle and lower quartiles respectively.  For the case shown in figure 4, the quartile values are $m_{\mathrm{Kep}}=12.0$, $11.5$ and $10.9$ respectively.

We may then calculate these quartile values for different exomoon masses, for a given star type.  Effectively, we are able to determine the minimum detectable exomoon mass as a function of magnitude for three different detection yields: i) 25\%, ii) 50\%, and iii) 75\%.  In figure 5, we show an example of one of these plots for an M2V star with a habitable-zone Saturn host planet.  We do not consider magnitudes brighter than $6^{\mathrm{th}}$ magnitude, since \emph{Kepler's} field-of-view has been selected to omit such stars. 

\begin{figure}
\begin{center}
\includegraphics[width=8.4 cm]{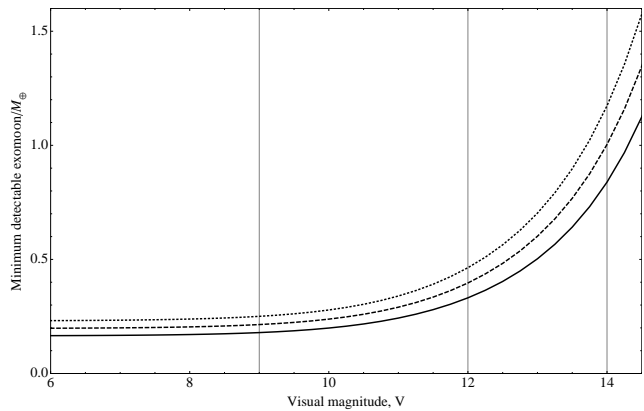}
\caption{\emph{Minimum detectable exomoon mass around a Saturn-like planet host in the habitable zone of an M0V star.  The top line is the 75\% catch-rate, the middle is 50\% and the lowest is for a 25\% exomoon catch-rate.  We also mark three key magnitudes of interest; 9, 12 and 14 magnitudes.}} \label{fig:fig5}
\end{center}
\end{figure}

\subsection{Minimum detectable habitable exomoon masses}

To complete the picture, we wish to see how these values differ for various star types.  We repeat the analysis performed above for M5V, M2V, M0V, K5V, K0V, G5V, G2V and G0V-type stars.  In figure 6, we plot the minimum detectable exomoon mass, for several contours of visual magnitude, as a function of stellar mass, in the 25\% detection-yield case.  We also show the upper limit on moon mass calculated from \citet{bar02} but this is only an approximation since we have assumed values of $Q_P$ and $k_{2p}$.

\begin{figure}
\begin{center}
\includegraphics[width=8.4 cm]{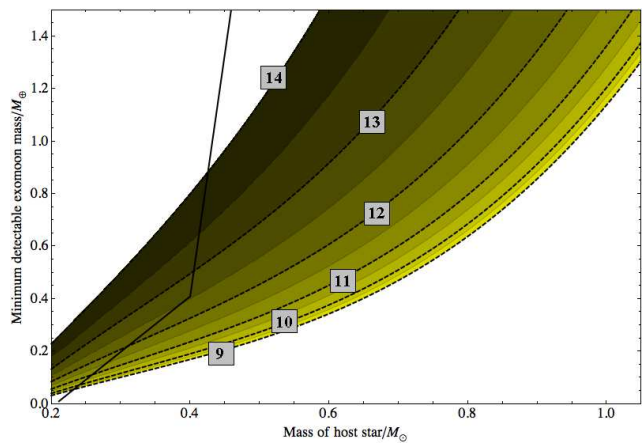}
\caption{\emph{Minimum detectable exomoon mass around a Saturn-like planet host in the habitable zone of various star types.  Contours show the different magnitude limits for a 25\% exomoon-catch-rate and overlaid is the mass stability limit (black, solid), found using \citet{bar02}.}} \label{fig:fig6}
\end{center}
\end{figure}

The $m_{\mathrm{Kep}}=12.5$ limit almost intersects the \citet{bar02} limit at $0.4 M_{\odot}$ and $\sim 0.4 M_{\oplus}$ for the exomoon corresponding to an M2V star $\sim 40$ pc away.  For comparison, within $10,000\, \mathrm{pc}^3$ (a spherical shell of radius 13.4 pc) there are $\sim 800$ main-sequence stars, $\sim 630$ of which are M-dwarfs (see \citet{led01}).

If we set a lower limit of 10 pc for our target star, then the lowest-mass habitable-zone exomoon which could be detected with KCP around a M5V star would be $0.09 M_{\oplus}$, which is above the maximum allowed stable mass limit of $0.008 M_{\oplus}$.  However, moving up to an M2V star we find a minimum detectable habitable exomoon mass of $0.18 M_{\oplus}$ which is below the stability limit of $0.41 M_{\oplus}$.  We therefore conclude that the minimum detectable habitable-zone exomoon mass with KCP is $\sim 0.2 M_{\oplus}$.

\subsection{1 $M_{\oplus}$ Limit}

For the 25\% detection yield case, we can convert the magnitude limit for detecting a 1 $M_{\oplus}$ habitable exomoon into a distance limit by making use of the absolute magnitudes for each star type.  The distance limit for detecting a 1 $M_{\oplus}$ habitable exomoon may then be compared to the limit found for a transiting 1 $R_{\oplus}$ habitable exoplanet.  In figure 7, we compare the two distance limits, which reveals that the distance limit for moons takes the same shape as the transit limit but is $\sim 1.6$ times lower.  This translates to a volume space diminished by a factor of $\sim 4$.

\begin{figure}
\begin{center}
\includegraphics[width=8.4 cm]{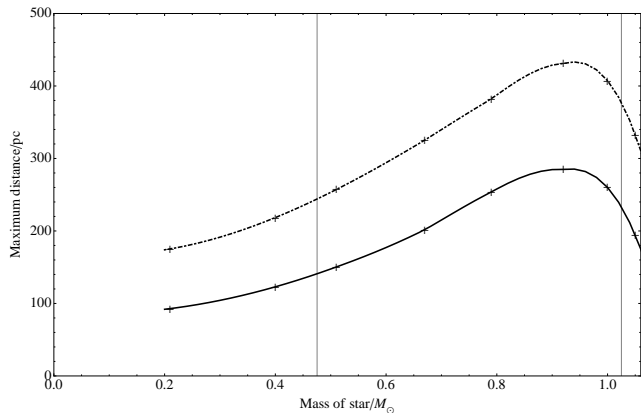}
\caption{\emph{Distance limit for detecting a 1 $M_{\oplus}$ habitable exomoon (solid) and a 1 $R_{\oplus}$ exoplanet (dot-dashed) as a function of stellar mass, for main sequence stars.  We assume a Saturn-like planet as the host for the exomoon case.  We also mark the lower stellar mass stability limit of $\sim 0.475$ $M_{\odot}$ and the upper mass limit of $\sim 1$ $M_{\odot}$ imposed by the requirement to observe $\geq 3$ transits.}} \label{fig:fig7}
\end{center}
\end{figure}

It is not our intention to estimate accurately how many observable stars might be contained within this volume, but one can deduce an order-of-magnitude estimate. Assuming that there are
about $10^5$ useable stars in \emph{Kepler}'s field of view, we estimate roughly that 25,000 stars would be within range for habitable-exomoon detection. Extrapolating \emph{Kepler}'s field of view to a band centred on the Galactic plane between $b=\pm15^\circ$, and simplistically assuming the same number of target stars per square degree, we obtain about $2\times 10^6$ stars that are accessible to KCP for a search for habitable-zone exomoons, if each star had a habitable-zone gas giant similar to Saturn.

\subsection{Detectability of an Earth-Moon analogue}

Finally, we consider the detectability of an Earth-Moon system.  Using the orbital parameters of the Sun-Earth-Moon system and assuming the most favourable configuration of $i=90^{\circ}$, we find the Earth would exhibit a TTV r.m.s. amplitude of 112.4 seconds and a TDV of 13.7 seconds.  The TDV is particularly low due to the large value of $\xi$ for the Moon.

In comparison, KCP of a $V=6$ star would yield timing errors of $\sigma_{t_c} = 389.6$ seconds and $\sigma_{T} = 779.2$ seconds.  Even for the most favourable case then, the TTV has a signal-to-noise of 0.3, which would be quite undetectable.  We therefore conclude that KCP cannot detect a Sun-Earth-Moon analogue through transit timing effects.

\section{Conclusions}

Our analysis finds that habitable exomoons are detectable up to $\sim 100$-$200$ pc away around early-M, K and later-G dwarf stars with the \emph{Kepler Mission} or photometry of equal quality (KCP).  This photometric quality should be sensitive down to $0.2 M_{\oplus}$ habitable exomoons in the idealised cases and could survey $\sim 10^6$ stars for $1 M_{\oplus}$ habitable exomoons, with a full Galactic Plane survey.  This number is around 25,000 stars for \emph{Kepler}'s field-of-view.

We find that Saturn-like planets are the ideal host candidates for detection due to their large radius to mass ratio.  Additionally, we find lower-mass exomoons may be found around M-dwarfs due to the closer habitable zone permitting a larger number of transits in a 4-year window.

The exciting prospect of discovering a habitable exomoon is well within the grasp of KCP and whether such worlds exist or whether they will be classed as truly habitable worlds are questions we can hope to answer in the coming years.  It is interesting to note that $\sim 0.2 M_{\oplus}$ is the minimum habitable mass for a planet or moon used by several other sources in the literature, including \citet{ray07} ($0.3 M_{\oplus}$) and \citet{wil97} ($0.12 M_{\oplus}$).  It has also been proposed by \citet{sch06} that the habitable zone could be extended for exomoons due to tidal heating and so our calculations may infact be an underestimate.

We find that \emph{Kepler} will be incapable of finding Ganymede-mass moons around Saturn-like planets.  \citet{can06} have suggested that the maximum moon mass which could form from a planetary debris disk is $2 \times 10^{-4}$ of that of the primary's and therefore, if this rule holds true, it unlikely KCP would detect any moons which formed around a planet in such a way.  However, moons above this mass limit are still dynamically stable (according to \citet{bar02}) and could have been captured by a planet or formed through an impact, for example like the formation of Triton and the Moon respectively.  Whether or not such objects are common is unknown but KCP could make the first in-depth search.

Our results suggest it is easier to detect an Earth-like exoplanet than an Earth-like exomoon around a gas giant.  However, we have no statistics to draw upon to estimate which of these scenarios is more common.  If a roughly equal number of both are discovered, it would indicate that the latter is more common due to the detection bias.

These results highlight the promising opportunity of making the first exomoon detection using the \emph{Kepler} telescope, or photometry of equivalent quality, especially the feasibility of detecting habitable-zone exomoons.  All-sky surveys focussing on bright M-dwarf stars would be ideally placed to search for habitable exomoons in greater depth and thus a telescope like \emph{TESS} could continue the search after the \emph{Kepler Mission} ends.

\section*{Acknowledgments}

DMK is supported by STFC, University College London and HOLMES ANR-06-BLAN-0416.  GC is supported by Ateneo Federato della Scienza e della Tecnologia - Universit\'a di Roma "La Sapienza", Collegio universitario `Don Nicola Mazza' and LLP-Erasmus Student Placement.  We extend special thanks to Bill Borucki for his help with estimating the noise properties of \emph{Kepler}.  Thanks to the \emph{Kepler Mission} technical team for inspiring this project.

\label{lastpage}


\begin{thebibliography}{99}
\bibitem[\protect\citeauthoryear{Agol et al.}{2005}]{ago05} Agol, E., Steffen, J., Sari, R \& Clarkson, W. 2005, MNRAS, 359, 567
\bibitem[\protect\citeauthoryear{Barnes \& O'Brien}{2002}]{bar02} Barnes, J. W. \& O'Brien, D. P., 2002, ApJ, 575, 1087
\bibitem[\protect\citeauthoryear{Basri et al.}{2005}]{bas05} Basri, G., Borucki, W. J. \& Koch, D., 2005, New Astronomy Rev., 49, 478
\bibitem[\protect\citeauthoryear{Batalha et al.}{2002}]{bat02} Batalha, N. M., Jenkins, J., Basri, G. S., Borucki, W.J., Koch, D.G., 2002, in Favata, F., Roxburgh, I. W., \& Galadi-Enriquez, D. (eds.), `Stellar Structure and Habitable Planet Finding' (ESA SP-485), p.35.
\bibitem[\protect\citeauthoryear{Butler et al.}{2002}]{but02} Butler, R. P. et al. 2002, ApJ, 578, 565
\bibitem[\protect\citeauthoryear{Borucki et al.}{2005}]{bor05} Borucki, W. J. et al., 2005, in M. Livio, K. Sahu \& J. Valenti (eds.), `A Decade of Extrasolar Planets around Normal Stars', (Cambridge University Press, Cambridge), p.36 (B05)
\bibitem[\protect\citeauthoryear{Cabrera \& Schneider}{2007}]{cab07} Cabrera, J. \& Schneider, J. 2007, A\&A, 464, 1133
\bibitem[\protect\citeauthoryear{Canup \& Ward}{2006}]{can06} Canup, R. M. \& Ward, W. R., 2006, Nature, 441, 834
\bibitem[\protect\citeauthoryear{Carter et al.}{2008}]{car08} Carter, J. A., Yee, J. C., Eastman, J., Gaudi, S. B. \& Winn, J. N. 2008, ApJ, 689, 499
\bibitem[\protect\citeauthoryear{Claret}{2004}]{cla04} Claret, A. 2004, A\&A, 428, 1001
\bibitem[\protect\citeauthoryear{Cox}{2000}]{cox00} Cox, A. N. (ed.), Allen's Astrophysical Quantities (4th edition) (Springer, Heidelberg), 2000.
\bibitem[\protect\citeauthoryear{Domingos et al.}{2006}]{dom06} Domingos, R. C., Winter, O. C. \& Yokoyama, T. 2006, MNRAS, 373, 1227
\bibitem[\protect\citeauthoryear{Dorren et al.}{1994}]{dor94}Dorren, J.D., Guinan, E.F., Dewarf, L.E. 1994, in J.-P. Caillault (ed). `Cool stars, stellar
systems, and the Sun', ASP Conf. Series, vol. 64, p.399. (Astronomical Society of the Pacific, San Francisco)
\bibitem[\protect\citeauthoryear{Han}{2008}]{han08} Han, C. 2008, ApJ, 684, 684
\bibitem[\protect\citeauthoryear{Holman \& Murray}{2005}]{hol05} Holman, M. J. \& Murray, N. W. 2005, Science, 307, 5713
\bibitem[\protect\citeauthoryear{Holman et al.}{2006}]{hol06} Holman, M. J. et al. 2006, ApJ, 652, 1715
\bibitem[\protect\citeauthoryear{Johnson et al.}{2009}]{joh09} Johnson, J. A., Winn, J. N., Cabrera, N. E. \& Carter, J. A. 2009, ApJ, 692, 100
\bibitem[\protect\citeauthoryear{Kipping}{2008}]{kip08} Kipping, D. M., 2008, MNRAS, 389, 1383
\bibitem[\protect\citeauthoryear{Kipping}{2009a}]{kipa09} Kipping, D. M., 2009a, MNRAS, 392, 181
\bibitem[\protect\citeauthoryear{Kipping}{2009b}]{kipb09} Kipping, D. M., 2009b, MNRAS, 396, 1797
\bibitem[\protect\citeauthoryear{Koch et al.}{2004}]{koc04} Koch D., 2004, in R. P. Norris and F. H. Stootman (eds.), IAU Symp. 213, `Bioastronomy 2002, Life Among the Stars' (Astronomical Society of the Pacific, San Francisco)
\bibitem[\protect\citeauthoryear{Koch et al.}{2007}]{koc07} Koch, D., Borucki, W., Basri, G. et al., 2007, in W.I. Hartkopf, E.F. Guinan \& P. Harmanec (eds.), `Binary Stars as Critical Tools \& Tests in Contemporary Astrophysics', Proc. IAU Symp. 240, p.236 (Cambridge University Press, Cambridge)
\bibitem[\protect\citeauthoryear{Ledrew}{2001}]{led01} Ledrew, G. 2001, JRASC, 95, 32L
\bibitem[\protect\citeauthoryear{Lewis et al.}{2008}]{lew08} Lewis, K. M., Sackett, P. D., Mardling, R. A, 2008, ApJ, 685, L153
\bibitem[\protect\citeauthoryear{Mandel \& Agol}{2002}]{man02} Mandel, K. \& Agol, E. 2002, ApJ, 580, L171
\bibitem[\protect\citeauthoryear{Metcalfe \& Charbonneau}{2003}]{met03} Metcalfe, T. S. \& Charbonneau, P. 2003, Journal of Computational Physics, 185, 176
\bibitem[\protect\citeauthoryear{Moutou et al.}{2006}]{mou06} Moutou, C., Pont, F., Bouchy, F \& Mayor, M. 2004, A\&A, 424, L31
\bibitem[\protect\citeauthoryear{Press et al.}{1992}]{pre92} Press W. H., Teukolsky S. A., Vetterling W. T., Flannery B. P. 1992, \emph{Numerical Recipes in FORTRAN77}, (Cambridge Univ. Press, Cambridge)
\bibitem[\protect\citeauthoryear{Raymond et al.}{2007}]{ray07} Raymond, S. N., Scalo, J. \& Meadows, V. S. 2007, ApJ, 669, 606
\bibitem[\protect\citeauthoryear{Sartoretti \& Schneider}{1999}]{sar99} Sartoretti, P. \& Schneider, J., 1999, A\&AS, 14, 550
\bibitem[\protect\citeauthoryear{Scharf}{2006}]{sch06} Scharf, C. A. 2006, ApJ, 648, 1196
\bibitem[\protect\citeauthoryear{Seager \& Mall\'{e}n-Ornelas}{2003}]{sea03} Seager, S., \& Mall\'{e}n-Ornelas, G. 2003, ApJ, 585, 1038
\bibitem[\protect\citeauthoryear{Simon et al.}{2009}]{sim09} Simon, A., Szab\'{o}, Gy. M. \& Szatm\'{a}ry, K. 2009, accepted by EM\&P
\bibitem[\protect\citeauthoryear{Szab\'{o}  et al.}{2006}]{sza06} Szab\'{o}, Gy. M., Szatm\'{a}ry, K., Div\'eki, Zs. \& Simon, A.,  2006, A\&A, 450, 395
\bibitem[\protect\citeauthoryear{Williams et al.}{1997}]{wil97} Williams, D. M., Kasting, J. F. \& Wade, R. A. 1997, Nature, 385, 235
\bibitem[\protect\citeauthoryear{Yee \& Gaudi}{2008}]{yee08} Yee, J. C. \& Gaudi, S. B., 2008, ApJ, 688, 616
\end{thebibliography}
\end{document}